\documentclass[10pt,reprint,superscriptaddress,aps,numbers,sort&compress]{revtex4-2}
\usepackage{times,amsmath,epsfig}
\usepackage{enumitem}
\usepackage{array}
\usepackage{graphicx}
\usepackage{float}
\usepackage{gensymb}
\usepackage{afterpage}
\usepackage{chemformula}
\usepackage{braket}
\usepackage{dcolumn}
\usepackage{bm}
\usepackage{hhline}
\usepackage{enumerate}
\usepackage{soul}
\usepackage{amsmath,amsthm,amssymb,mathrsfs} 

\usepackage[LGRgreek]{mathastext}
\newcolumntype{L}[1]{>{\raggedright\let\newline\\\arraybackslash\hspace{0pt}}m{#1}}
\newcolumntype{C}[1]{>{\centering\let\newline\\\arraybackslash\hspace{0pt}}m{#1}}
\newcolumntype{R}[1]{>{\raggedleft\let\newline\\\arraybackslash\hspace{0pt}}m{#1}}

\begin{document}

\title{Coherent multi-dimensional widefield microscopy}

\author{Mohammadjavad Azarm}
\email[]{mohammadjaval.azarm@unicatt.it}
\affiliation{Department of Mathematics and Physics, Università Cattolica del Sacro Cuore, Brescia I-25133, Italy}
\affiliation{ILAMP (Interdisciplinary Laboratories for Advanced
Materials Physics), Università Cattolica del Sacro Cuore, Brescia I-25133, Italy}
\affiliation{Department of Physics and Astronomy, KU Leuven, B-3001 Leuven, Belgium}

\author{Rizwan Asif}
\affiliation{Department of Mathematics and Physics, Università Cattolica del Sacro Cuore, Brescia I-25133, Italy}
\affiliation{ILAMP (Interdisciplinary Laboratories for Advanced
Materials Physics), Università Cattolica del Sacro Cuore, Brescia I-25133, Italy}
\affiliation{Department of Biosystems, KU Leuven, B-3001 Leuven, Belgium}
\author{Alessandra Milloch}
\affiliation{Department of Mathematics and Physics, Università Cattolica del Sacro Cuore, Brescia I-25133, Italy}
\affiliation{ILAMP (Interdisciplinary Laboratories for Advanced
Materials Physics), Università Cattolica del Sacro Cuore, Brescia I-25133, Italy}
\affiliation{Department of Physics and Astronomy, KU Leuven, B-3001 Leuven, Belgium}
\author{Francesco Proietto}
\affiliation{Department of Mathematics and Physics, Università Cattolica del Sacro Cuore, Brescia I-25133, Italy}
\affiliation{ILAMP (Interdisciplinary Laboratories for Advanced
Materials Physics), Università Cattolica del Sacro Cuore, Brescia I-25133, Italy}
\affiliation{Department of Physics and Astronomy, KU Leuven, B-3001 Leuven, Belgium}
\author{Donna Datta}
\affiliation{Elmore Family School of Electrical and Computer Engineering, Purdue University, 465
Northwestern Ave, West Lafayette, IN, USA}
\author{Ambrine Lanseur}
\affiliation{Université Claude Bernard Lyon 1, Villeurbanne, France}

\author{Sreyan Raha}
\author{Filippo Fabbri}
\author{Federica Bianco}
\affiliation{NEST Laboratory, Istituto Nanoscienze-CNR, Piazza S. Silvestro 12, I-56127 Pisa, Italy}

\author{Fabrizio Preda}
\affiliation{NIREOS SRL, Via G. Durando 39, 20158 Milano, Italy}

\author{Antonio Perri}
\affiliation{NIREOS SRL, Via G. Durando 39, 20158 Milano, Italy}

\author{Suman Chatterjee}
\affiliation{Quantum Matter Institute, University of British Columbia, Vancouver, British Columbia, V6T 1Z4 Canada}

\author{Ziliang Ye}
\affiliation{Department of Physics \& Astronomy, University of British Columbia, Vancouver, British Columbia, V6T 1Z1 Canada}
\affiliation{Quantum Matter Institute, University of British Columbia, Vancouver, British Columbia, V6T 1Z4 Canada}

\author{Giulio Cerullo}
\affiliation{Department of Physics, Politecnico di Milano, Piazza Leonardo da Vinci 32, I-20133 Milano, Italy}
\affiliation{IFN-CNR, Piazza Leonardo da Vinci 32, I-20133, Milano, Italy}

\author{Stefania Pagliara}
\affiliation{Department of Mathematics and Physics, Università Cattolica del Sacro Cuore, Brescia I-25133, Italy}
\affiliation{ILAMP (Interdisciplinary Laboratories for Advanced
Materials Physics), Università Cattolica del Sacro Cuore, Brescia I-25133, Italy}

\author{Gabriele Ferrini}
\affiliation{Department of Mathematics and Physics, Università Cattolica del Sacro Cuore, Brescia I-25133, Italy}
\affiliation{ILAMP (Interdisciplinary Laboratories for Advanced
Materials Physics), Università Cattolica del Sacro Cuore, Brescia I-25133, Italy}

\author{Claudio Giannetti}
\email[]{claudio.giannetti@unicatt.it}
\affiliation{Department of Mathematics and Physics, Università Cattolica del Sacro Cuore, Brescia I-25133, Italy}
\affiliation{ILAMP (Interdisciplinary Laboratories for Advanced
Materials Physics), Università Cattolica del Sacro Cuore, Brescia I-25133, Italy}
\affiliation{CNR-INO (National Institute of Optics), via Branze 45, 25123 Brescia, Italy}

\begin{abstract}
Understanding how electronic excitations evolve across space and time is essential for revealing the microscopic processes underlying quantum and optoelectronic materials. However, existing approaches cannot simultaneously resolve ultrafast coherent dynamics with microscopic spatial information: pump–probe microscopy lacks access to quantum coherence, while two-dimensional electronic spectroscopy (2DES) requires slow point-by-point scanning for spatial resolution. Here we introduce a widefield two-dimensional electronic spectroscopy microscope (2DESM) combining multidimensional coherent spectroscopy with widefield imaging, enabling simultaneous femtosecond temporal and micrometer spatial resolution across a broadband spectral range. Applying 2DESM to hBN-encapsulated WSe$_2$, we directly visualize spatial variations in exciton coherence and relaxation associated with the local environment, establishing 2DESM as a powerful platform for probing many-body interactions, energy transport, and structure–function relationships in low-dimensional materials and devices.
\end{abstract}

\maketitle

\section{INTRODUCTION}
Recent advances in the synthesis of low-dimensional quantum materials \cite{wang2025electronic}, van der Waals heterostructures \cite{Liu2016,li2020general}, and nanodot artificial solids \cite{Lee2011,Protesescu2015,Raino2018,Milloch2023} have underscored the need for advanced spectroscopic techniques capable of studying the fundamental ultrafast electronic processes with both spectral and spatial resolution \cite{liu2024super}. Transition metal dichalcogenides (TMDs) \cite{Manzeli2017,Wang2018} represent a paradigmatic system in which well defined optical excitons can experience spatially dependent interlayer and intralayer potentials in heterostructures \cite{rivera2018interlayer,Jin2019,Tran2019,Naik2022,du2024new} and Moiré potentials in twisted bilayers \cite{Seyler2019,Zhang2020,choi2020moire}. Under strong excitation, TMDs are test bed for spatially inhomogeneous excitonic Mott transitions \cite{Chernikov2015,Guerci2019} as well as for the formation of many-body complexes such as biexcitons and trions \cite{wang2014valley}. Furthermore, the reduced dimensionality and structural heterogeneity \cite{choi2020moire} boost microscale phenomena including exciton diffusion \cite{lopez2023tip,choi2020moire}, charge transfer \cite{krause2021microscopic, fabbri2025interplay}, and edge-localization \cite{sui2022ultrafast}. 


So far, conventional approaches to studying non-equilibrium dynamics in these systems have predominantly relied on ultrafast pump–probe spectro-microscopy. \cite{Domke2012,Schnedermann2016,piland2019high}, which consists in the combination of broadband pump-probe optical spectroscopy with microscopy. This approach is effective in probing the space-dependent population dynamics \cite{chen2021ultrafast,Cui2014,Zhang2022,Genco2023,Faini2025}, but it is inherently insensitive to those processes that involve the quantum-coherent evolution of the macroscopic polarization. Examples are the ultrafast quantum coherent evolution of the electronic wave-packets \cite{trovatello2020ultrafast}, the inhomogeneous broadening associated to local disorder \cite{moody2015intrinsic}, and the coherent coupling between different electronic states \cite{purz2021coherent}, which impact both the fundamental understanding and the device performances. 

Multidimensional electronic spectroscopy \cite{cundiff2013optical,smallwood2018multidimensional}, based on the excitation with multiple phase-coherent light pulses, has emerged as the go-to technique to overcome many of the intrinsic limitations of traditional pump-probe techniques. In the case of two phase-coherent excitation pulses, Two-Dimensional Electronic Spectroscopy (2DES) allows to correlate excitation and detection frequencies while simultaneously preserving high spectral and temporal resolution, thereby allowing the direct observation of coherent dynamics, energy transfer pathways, and many-body interactions with femtosecond precision \cite{kim2009two,collini20212d,czech2015measurement,maiuri2019ultrafast,biswas2022coherent,timmer2024ultrafast,bressan2024two}. 2DES has proven especially powerful in disentangling homogeneous and inhomogeneous broadening \cite{kuznetsova2007determination,moody2015intrinsic}, providing access to the intrinsic linewidth of optical excitations even in spectrally disordered systems \cite{biswas2022coherent}.  It also enables the identification of electronic couplings, energy-transfer pathways, and coherent interactions between states \cite{cundiff2013optical}. \\
Efforts to complement multi-dimensional spectroscopies with spatial resolution have so far primarily relied on point scanning approaches employing tightly focusing optics \cite{Jakubczyk2016, Jakubczyk2018, Jakubczyk2019, liu2025nonlocal} or on tip-enhanced nanoimaging \cite{Luo2023}. Here, we present a widefield microscope integrating the capabilities of conventional 2DES, which we refer to as 2DESM. This setup preserves high temporal resolution (tens of femtoseconds) and broad spectral bandwidth, and enables time resolved spectral imaging that captures ultrafast coherent electronic dynamics with micrometer-scale resolution.
We apply 2DESM to investigate the ultrafast coherent excitonic dynamics in a bilayer $WSe_2$ flake encapsulated by hBN, which provides a clean, strain-minimized environment preserving the intrinsic optical properties of the material. We demonstrate the possibility of measuring space-dependent 2DES maps and extract direct information about the local decoherence dynamics and the disorder-induced broadening of excitonic lines in $WSe_2$.
More in general, this novel integrated platform paves the way for studying quantum-coherent wave-packet propagation, space diffusion of excitons and many-body complexes (e.g. biexcitons, trions), and non-local coherent energy exchange between different electronic excitations, with broad application to low-dimensional materials, van der Waals heterostructures \cite{Regan2022}, and micrometer-sized optoeletronic devices.
\begin{figure*}[t] 
\centering
\includegraphics[width=1\linewidth]{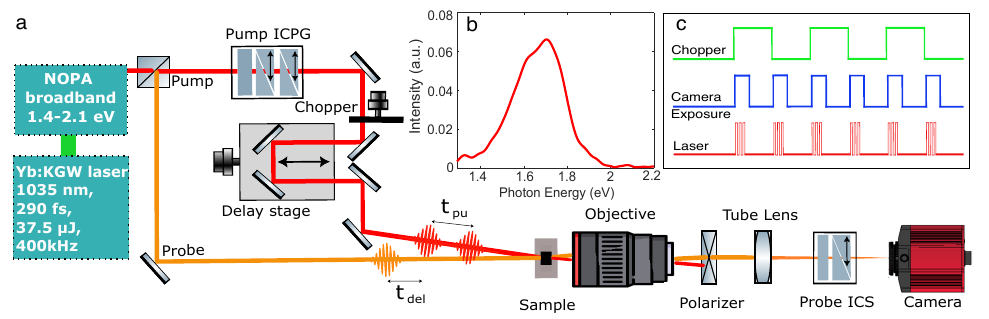}
\caption{\label{fig:setup}\textbf{(a) Schematic of the 2DESM setup} – The light source is a home-built non-collinear optical parametric amplifier (NOPA). The NOPA produces broadband light tunable from 1.4 to 2.1 eV, generating spectrally identical pump and probe pulses. A pair of pump pulses is generated by the Interferometric Collinear Pulse Generator (ICPG), which creates two phase-locked replicas with a controllable time delay $t_{pu}$. The pump–probe delay $t_{del}$ is introduced via a mechanical delay stage, and the chopper provides the timing reference for synchronizing acquisition. After interaction with the sample, a wide-field objective and tube lens image the transmitted probe onto the camera, while a polarizer suppresses residual pump scattering to enhance the SNR. The Interferometric 
Collinear Spectrometer (ICS) is placed after the tube lens and generates two temporally delayed probe pulses with delay $t_{pr}$. The black double arrow in both ICPG and ICS indicates
the wedge translation direction. \textbf{(b) NOPA spectrum} – The NOPA output is centered at 1.70 eV, covering a range from 1.41 to 1.82 eV.  \textbf{(c) Timing diagram of the triggering system} – The chopper serves as the master timing reference, while the camera acquires two frames per chopper duty cycle. Each camera acquisition triggers the laser to emit a defined number of pulses, set by the camera exposure time. In this scheme, the chopper blocks the pump beam for one of the two frames, yielding one frame with the pump on and one with the pump blocked.}
\end{figure*}
\section{2DES microscopy}
Our technique builds on 2DES, a non-linear optical spectroscopy grounded in Four-Wave Mixing (FWM) \cite{Mukamel,cundiff2013optical}. In this scheme, a third-order polarization $P^{(3)}(t)$ is generated through the interaction of three ultrafast laser pulses, i.e. two phase-coherent pump pulses and a third delayed probe, with the sample. By simultaneously resolving excitation and detection frequencies, 2DES extends the capabilities of conventional one-dimensional time-resolved techniques, such as broadband pump–probe spectroscopy, generating two-dimensional correlation maps that reveal couplings and dynamics inaccessible to standard approaches \cite{piland2019high}. 
The standard 2DES signal is given by:
\begin{equation}
\label{eq.1 2DES Signal}
    S( \omega_{pu}, \omega_{pr},t_{del})
\end{equation}
where $\omega_{pu}$ denotes the excitation (pump) frequency, $\omega_{pr}$ the detection (probe) frequency, and $t_{del}$ the time delay between the pump and probe pulses. To measure $S( \omega_{pu}, \omega_{pr},t_{del})$, a pair of phase-locked pump pulses 
spaced by a controllable time delay $t_{pu}$ is used to create a coherent superposition of quantum states.  Scanning $t_{pu}$ and performing a Fourier transform (FT) yield the excitation frequency axis $\omega_{pu}$ of the 2D spectrum. After interacting with the two pump pulses, the system is probed by the third pulse.
The emitted signal is 
spectrally resolved by an interferometric spectrometer, providing resolution along the detection frequency axis ($\omega_{pr}$).
We note that, although in conventional 2DES notation the coherence and waiting times are denoted by $t_1$ and $t_2$, respectively\cite{jonas2003two}, in this work they will be indicated as $t_{pu}$ and $t_{del}$ for sake of simplicity in describing the experimental setup and data-acquisition procedure.
The implementation of 2DES in a partially collinear geometry \cite{Hamm_Zanni_2011,Fuller2015} simplifies phase matching, enhances alignment stability, and supports transmission-based detection, features that make it particularly well suited for studying complex condensed matter systems \cite{cundiff2013optical,Giannetti2016,maiuri2019ultrafast}.

Here we extend 2DES into the spatial domain by collecting the probe pulse, after the interaction of the pulses with the sample, via an optical widefield microscope. In this way it is possible to perform an optical imaging of the sample with high spatial resolution, without losing the temporal and multidimensional spectral information. Each set of acquired data corresponds to a specific pair of excitation and detection frequencies, ($\omega_{pu}$, $\omega_{pr}$), that evolve as a function of time delay $t_{del}$ (Eq.\ref{eq.1 2DES Signal}), while spatial coordinates ($x, y$) define the position within the field of view. The full data volume  is then expressed as
\begin{equation}
\label{eq.2 2DES map}
    S(x,y;\omega_{pu}, \omega_{pr}; t_{del})
\end{equation}
and contains spatially resolved 2DES maps.
\subsection{Optical setup}
In Figure \ref{fig:setup}a we show a sketch of the 2DESM setup. The light source is a home-built non-collinear optical parametric amplifier (NOPA), pumped by an amplified Yb:KGW laser (Pharos by Light Conversion) delivering 37.5 $\mu J$ pulses at 400 kHz repetition rate. The NOPA generates broadband pulses, tunable from 1.4 to 2.1 eV. Figure \ref{fig:setup}b shows the typical NOPA spectrum, centered at 1.70 eV and covering the $1.41-1.82$ eV spectral range. The NOPA output beam is split to feed both the pump and probe lines. The pump arm passes through an Interferometric Collinear Pulse Generator (ICPG) (GEMINI 2D by
NIREOS). The ICPG is based on the Translating-Wedge-Based Identical Pulses eNcoding System (TWINS) scheme, as introduced in Ref. \citenum{brida2012phase}, in which birefringent elements introduce a precisely controlled optical delay between two temporally separated replicas of the pulse traveling along a common optical path. The ICPG generates two pulses at variable time delay $t_{pu}$, with phase stability
of the order of 1/1000 of the optical cycle.
The pump-probe time delay $t_{del}$ is controlled by the mechanical delay stage, which introduces an additional delay by adjusting the pump path with respect to the probe optical path. Due to the pulse stretching caused by propagation through optical components, such as polarizers, waveplates, and crystals inside the ICPG, two distinct pulse compressors based on pairs of chirped mirrors are employed for the pump and probe beams, respectively. Subsequently, the beams are focused onto the sample using concave mirrors with focal lengths of 40 cm for the pump and 20 cm for the probe, resulting in spot diameters (FWHM) of $\sim 190~\mu\text{m}$ and $\sim 100~\mu\text{m}$ for the pump and probe beams, respectively.
To collect the probe beam after its interaction with the sample excited by the pump pulses, we use a wide-field objective with 20X magnification, 20 mm working distance, and 0.40 numerical aperture. An infinity-corrected tube lens with 200 mm focal length and 151.8 mm working distance, operating in the $1.2–1.9 $ eV spectral range, images the sample onto the camera used for detection. 
A scientific high-speed camera (CS135MUN by Thorlabs) is used for rapid data acquisition synchronized with the laser source emission. 
The camera features a 1280$\times$1024 pixel detector array ($\sim1.3$ MP), with a pixel dimension of 4.8$\times$4.8 $\mu m^2$. By selectively cropping regions from the full frame, we are able to increase imaging speed up to 3469 frames per second with a minimum exposure time of 0.1 ms. \\

To extend 2DESM measurements to cryogenic temperatures, a custom cryogenic chamber was designed and integrated into a closed-cycle cryostat system. The chamber is equipped with a 1.5 mm-thick optical entrance window; the residual dispersion introduced by this element is pre-compensated by the chirped-mirror compressors. A 200 $\mu$m-thick optical exit window, placed between the sample and the microscope objective, allows the objective to operate at its nominal working distance, preserving both the numerical aperture and the spatial resolution of the imaging system while minimizing optical aberrations. Low-temperature operation of the 2DESM platform is demonstrated in Section \ref{sec.Temp}, where measurements on WSe$_2$ at 20 K are presented. 
\begin{figure*}[t] 
\centering
\includegraphics[width=1\linewidth]{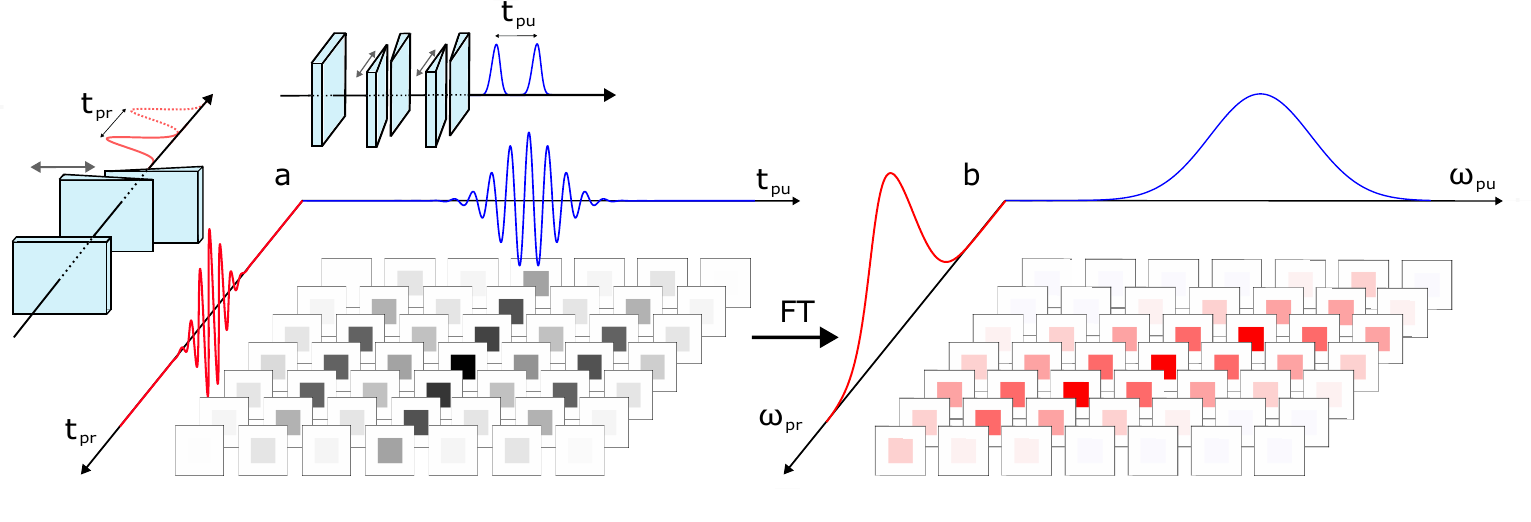}
\caption{\label{fig:2Ddata} 
\textbf{(a) Time-domain hypercube} – Schematics of the construction of the $S(x,y;t_{pu},t_{pr},t_{del})$ time-domain hypercube at the fixed pump-probe delay $t_{del}$. Each image, represented in the black squares, corresponds to a specific ($t_{pu}$,$t_{pr}$) coordinate. The top sketches illustrate the ICPG and ICS devices with black double arrows indicating
the direction of translating wedges to control the $t_{pu}$ and $t_{pr}$ coordinates. The red and blue interferograms represent the typical signals obtained for fixed ($x$,$y$) spatial coordinates upon scanning the $t_{pu}$ and $t_{pr}$ axes.
\textbf{(b) Frequency-domain hypercube} – A two-dimensional FT is applied for each image pixel, thus converting the time-domain $S(x,y;t_{pu},t_{pr},t_{del})$ hypercube into the frequency-domain hypercube $S(x,y;\omega_{pu},\omega_{pr},t_{del})$.}
\end{figure*}
\subsection{Hyperspectral Data Acquisition}
The synchronization between image acquisition and light pulses is essential to improve 
the signal to noise ratio (SNR) and extract the weak difference between pumped and unpumped images, as shown in the timing diagram of Figure \ref{fig:setup}c. A chopper wheel, positioned on the pump path, serves as the master timing reference (clock), triggering the camera to acquire two frames per cycle, corresponding to the pump pulse on and off, respectively. The clock signal (green line in Fig. \ref{fig:setup}c) is used to generate synchronized digital triggers for both the laser emission and the camera acquisition, thus ensuring that a controllable number of laser pulses is emitted within the camera’s exposure window. 
The number of laser pulses per cycle and individual delay offsets are tunable parameters which can be adapted and optimized depending on the various experimental conditions. 
For each chopper cycle, we acquire two distinct image streams ($I_{i} (x,y)$) corresponding to pumped ($i$: even) and unpumped ($i$: odd) signals. At 1.5 kHz chopping rate, the camera records $\simeq$3000 frames per second. 
To compensate intensity fluctuations and enhance the SNR, each image $I_{i} (x,y)$ is normalized, frame by frame, to its background intensity, $\overline{I}_{i}$. This is achieved by selecting a sub-region far enough from the sample, where no pump-induced signal is expected, and integrating the total intensity detected by the camera within selected region. This high speed normalization ensures consistency across frames by accounting for frame-to-frame variations in illumination and detector response \cite{hormann2024self,wu2025self}. 
For each value of $t_{pu}$ and $t_{del}$ we then compute the mean pumped ($\overline{T}_{pumped}$) and unpumped ($\overline{T}_{unpumped}$) transmitted intensity signals by averaging over $N$ acquired frames as: 
\begin{equation}
\label{eq:T_pumped}
    \overline{T}_{pumped} =\frac{1}{N} \sum^{N}_{\substack{i=1 \\ \text{even}}} \frac{I_{i} (x,y;t_{pu},t_{del})}{\overline{I}_{i}}
\end{equation}
and
\begin{equation}
\label{eq:T_unpumped}
    \overline{T}_{unpumped} =\frac{1}{N} \sum^{N}_{\substack{i=1 \\ \text{odd}}} \frac{I_{i} (x,y;t_{pu},t_{del})}{\overline{I}_{i}}.
\end{equation}
The relative transmittivity variation signal for fixed $t_{pu}$ and $t_{del}$ is obtained as:
\begin{equation}
    {\Delta T}(x,y;t_{pu},t_{del}) =  {\overline{T}_{pumped}-\overline{T}_{unpumped}}
    \label{eq:DT_T}.
\end{equation}
Within this imaging scheme, frequency resolution along the probe axis of the 2D spectrum is obtained by means of an Interferometric Collinear Spectrometer (ICS) (GEMINI by NIREOS), which is a simplified version of the ICPG previously introduced. Two birefringent wedges introduce a precisely controlled optical delay $t_{pr}$ \cite{brida2012phase} between two replicas of the probe light traveling along a common optical path from the tube lens to the camera (see Figure \ref{fig:setup}a) \cite{Candeo2019}. Application of FT to the modulated patterns returns the spectral content of the probe beam while preserving spatial resolution.

\begin{figure}
\centering
\includegraphics[width=1\linewidth]{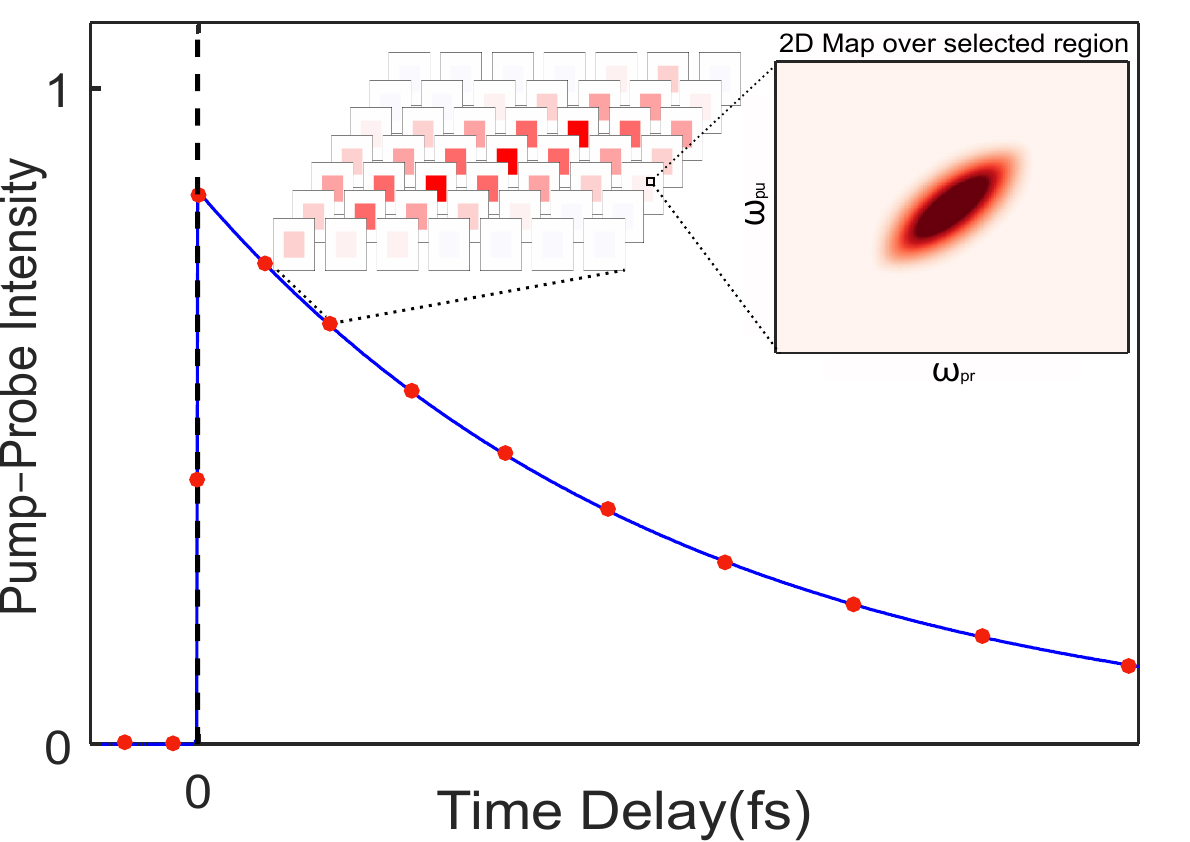}
\caption{\label{fig:2dmap} \textbf{Ultrafast 2DESM temporal dynamics} – Illustration of the full 2DESM acquisition process. For each pump-probe delay $t_{del}$ a full frequency-domain hypercube $S(x,y;\omega_{pu},\omega_{pr},t_{del})$ is acquired. For each spatial coordinate ($x$,$y$) a time-dependent 2DES spectrum,as reported in the top right panel, is obtained.}
\end{figure}

Overall, the coordinated action of ICPG on the pump beam, which controls $t_{pu}$, and ICS on the probe, which controls $t_{pr}$, enables to perform 2DES spectroscopy combined with widefield microscopy (2DESM). Figure \ref{fig:2Ddata} illustrates the overall acquisition process. At a fixed pump–probe delay $t_{del}$, a $\Delta T(x,y;t_{pu},t_{del})$ image (see Eq. \ref{eq:DT_T}) is recorded for each step of $t_{pu}$. The scanning across multiple values of $t_{pu}$ forms a data stack. A second scan is performed along the  $t_{pr}$ dimension, thus producing a multidimensional dataset (hypercube), i.e. $\Delta T(x,y;t_{pu},t_{pr},t_{del})$. A two-dimensional FT is then applied to extract the spectral information along both excitation and detection axes. The resulting two-dimensional signal  is obtained by:
\begin{equation}
\label{eq:signal}
\begin{aligned}
S(x,y;\omega_{pu},\omega_{pr},t_{del}) 
&= \int \int  {\Delta T}(x,y;t_{pu},t_{pr},t_{del}) \\
&\quad \times e^{-i \omega_{pu} t_{pu}} \, e^{-i \omega_{pr} t_{pr}} 
\, dt_{pu} \, dt_{pr}.
\end{aligned}
\end{equation}
Finally, to provide a calibrated and quantitative response, the spectrum at each pixel is normalized to the probe spectrum, correcting for spectral non-uniformities and referencing the signal to its equilibrium value. This procedure is repeated for each pixel of the image, corresponding to the coordinate ($x$,$y$). As illustrated in Figure \ref{fig:2Ddata}, this dual-scanning strategy yields, at a fixed $t_{del}$, a multidimensional hyperspectral datacube, with both excitation ($\omega_{pu}$) and detection ($\omega_{pr}$) spectral dimensions derived from the temporal delays controlled by the ICPG and ICS. 2DESM thus enables the multidimensional investigation of ultrafast coherent excitation dynamics in real materials and devices with high spatial resolution. As illustrated in Figure \ref{fig:2dmap}, at a selected pump–probe delay ($t_{del}$), a hyperspectral $S(x,y;\omega_{pu},\omega_{pr},t_{del})$ data cube is acquired. The top-right panel of Figure \ref{fig:2dmap}, displays a typical 2DES map that can be extracted at a specific ($x$,$y$) point. In the sketched example, the signal, elongated along the image diagonal represents the pump induced variation of an excitonic resonance, whose decoherence time and inhomogeneous broadening can be extracted from the analysis of the diagonal and off-diagonal line profiles, as it will be extensively discussed in Section \ref{sec:WSe2}.
\subsection{2DESM resolution}
\label{sec:resolution}
\begin{figure}
\centering
\includegraphics[scale=0.45]{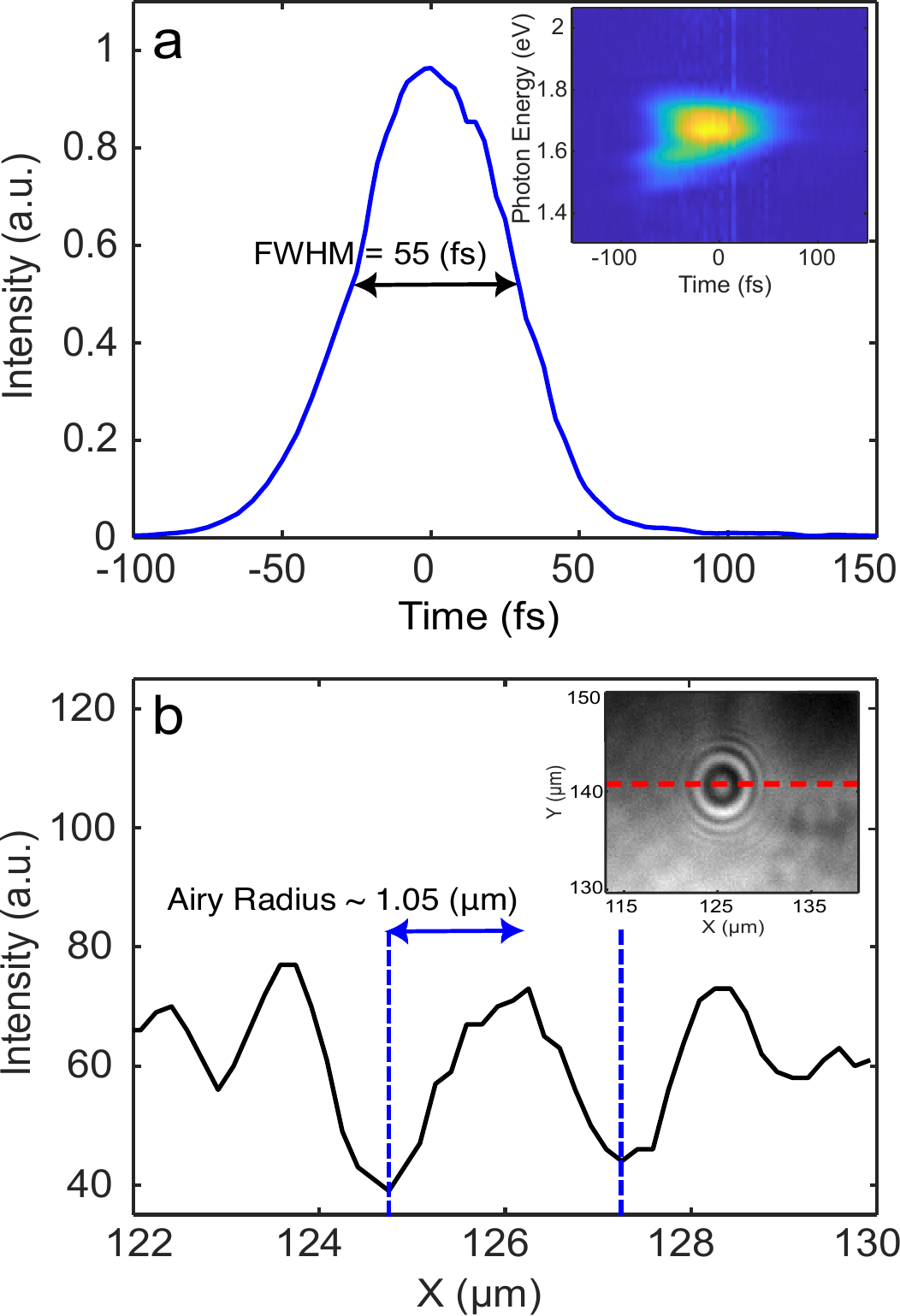}
\caption{\label{fig:resolution} 
\textbf{(a) Pulse duration} – The main panel displays the temporal profile obtained by integrating the PG-FROG signal (shown in the inset). \textbf{\textbf{(b) Spatial resolution}} – The main panel displays the line profile extracted from the diffraction pattern of a point-like defect. The spatial resolution of the system is estimated as half of the distance between the first two minima of the Airy disk. The diffraction pattern from the point-like defect is shown in the top-right panel. The red dashed line indicates the line along which the intensity profile reported in the main panel is taken.}
\end{figure}
A comprehensive assessment of the 2DESM setup performance was conducted by characterizing its temporal, spatial, and spectral resolution. 
Temporal resolution was evaluated using polarization-gated frequency-resolved optical gating (PG-FROG) \cite{trebino1997measuring}. Figure \ref{fig:resolution}a presents the cross-correlation signal between pump and probe pulses, at the sample position, obtained by integrating the PG-FROG trace along the wavelength axis. The inset of Figure \ref{fig:resolution}a displays the full PG-FROG trace, representing the cross-correlation as a function of time delay and wavelength. The measured full width at half maximum (FWHM) of the cross-correlation signal is 55 fs. Assuming identical Gaussian temporal profiles for both pulses, we estimated an individual pulse duration of approximately 38 fs on the sample. This value defines the temporal resolution limit of our system, which is not altered by the optical components (objective, tube lens, ICS) that collect the probe light after the non-linear interaction.

Spatial resolution was determined by imaging a sub-micrometer defect and analyzing its Airy disk diffraction pattern, as shown in Figure \ref{fig:resolution}b. Considering half of the distance between the first two intensity minima, the spatial resolution results to be approximately 1.05 µm. The inset of Figure  \ref{fig:resolution}b shows the image of the point-like defect, highlighting the resulting diffraction pattern under laser illumination.

In the TWINS scheme, the spectral resolution is controlled by the maximum temporal delay introduced via the translation of birefringent wedges \cite{perri2019hyperspectral}. This resolution can be quantitatively estimated using the relation:
\begin{equation}
\Delta\nu = \frac{c}{\Delta n  L  \tan(\alpha)} 
\end{equation}
where $c$ is the speed of light, $\Delta n$ represents the difference between the refractive indices of the ordinary and extraordinary polarization components, and $L$ is the total displacement range of the wedge with the apex angle $\alpha$. Since the spectral resolution is inversely proportional to the maximum delay, increasing the displacement range  $L$ enhances the resolution. To strike a balance between spectral resolution and acquisition speed, the wedge displacement range is selected as $L$=1.2 mm. Using birefringent wedges of BBO material ($\Delta$n=-0.1152 at 800 nm\cite{eimerl1987optical}) with an apex angle of $\alpha=10^{\circ}$, the spectral resolution was determined to be $\Delta\lambda$=22 nm at 800 nm central wavelength, corresponding to an energy resolution of 50 meV at 1.5 eV.
\section{2DESM on Bilayer WS\lowercase{e}$_2$}
\label{sec:WSe2}
TMDs constitute the ideal test bench for the 2DESM technique introduced in this work. The preparation of monolayer or few-layer samples via mechanical exfoliation typically yields micrometer-sized flakes, necessitating microscopy-based approaches. In terms of optical properties, the bi-dimensional nature of TMDs atomic layers strongly suppresses the bulk Coulomb screening thus giving rise to well defined excitons with a large binding energy as compared to bulk semiconductors \cite{Wang2018}. More specifically, a band gap of 2.53 eV and a binding energy $E^A_b$=0.79 eV have been reported  at $T$=4 K for the $A$-exciton in WSe$_2$ monolayers deposited on SiO$_2$  \cite{Hanbicki2015}. The main excitonic line thus corresponds to a photon energy $E^A_{exc}$=1.74 eV. Considering the typical band-edge redshift of semiconductors associated to electron-phonon coupling \cite{ODonnell1991}, at $T$=300 K the $A$-exciton energy decreases down to $E^A_{exc}\simeq$1.65 eV  \cite{Hanbicki2015}, which lies in the center of the pulse bandwidth (see Figure \ref{fig:setup}b). 

\begin{figure}[t] 
\centering
\includegraphics[width=1\linewidth]{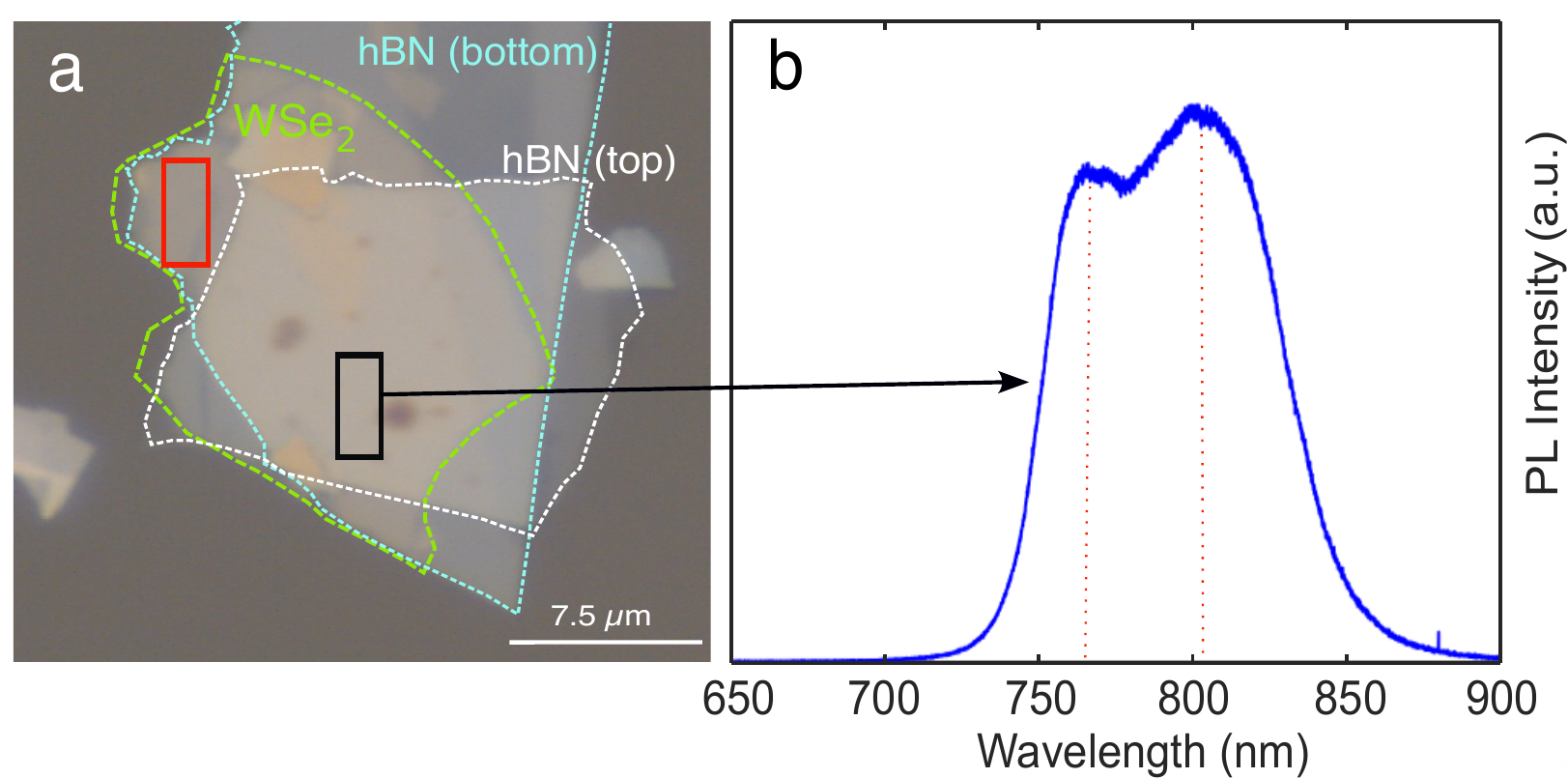}
\caption{\label{fig:sample}\textbf{(a) Bilayer WSe$_2$} – An optical image of the sample, taken with a 50x objective, is shown. The bilayer WSe$_2$ flake, indicated by the green contour, was mechanically exfoliated and deposited on a SiO$_2$ substrate covered by a hBN layer (light blue contour). The top side of WSe$_2$ has been protected by an additional hBN overlayer, indicated by the white contour. The black and red rectangles indicated the two regions that are taken as representative of fully encapsulated and unprotected WSe$_2$ bilayer.
\textbf{(b)} \textbf{Photoluminescence} – Photoluminescence measurements from the black rectangular area of panel (a) is shown. The spectrum displays two distinct peaks at 770 nm and 810 nm (see dashed red lines), consistent with emission from bilayer WSe$_2$ \cite{Zhao2013,Tonndorf2013}. }
\end{figure}

Here we apply 2DESM to investigate a bilayer WSe$_2$ flake placed on a transparent glass substrate, as shown in Figure \ref{fig:sample}a. 
WSe$_2$ (green dashed contour) has been encapsulated by bottom (light blue dashed contour) and top (white dashed contour) layers of hBN. The sample area measures approximately 15$\times$15 $\mu m^2$ (green dashed line), and contains two distinct regions: 1) approximately at the center of the image, WSe$_2$ is encapsulated by both top and bottom hBN layers. The black rectangular area is taken as representative of fully encapsulated WSe$_2$; 2) the top-left region of the sample is not covered by the top hBN overlayer. The red rectangular area is taken as representative of unprotected WSe$_2$.

To characterize the sample, photoluminescence (PL) measurements were conducted.  As shown in Figure \ref{fig:sample}b the PL spectrum from the central area of the sample exhibits a weak emission peak near 770 nm (1.61 eV) and a second peak near 810 nm (1.53 eV), characteristic of indirect excitonic recombination in bilayer WSe$_2$. The presence of this secondary peak and the reduced intensity indicates interlayer coupling effects that modify the electronic band structure, confirming the bilayer nature of the WSe$_2$ flake \cite{Zhao2013,Tonndorf2013}.

\begin{figure*}[t]
\centering
\includegraphics[width=1\linewidth]{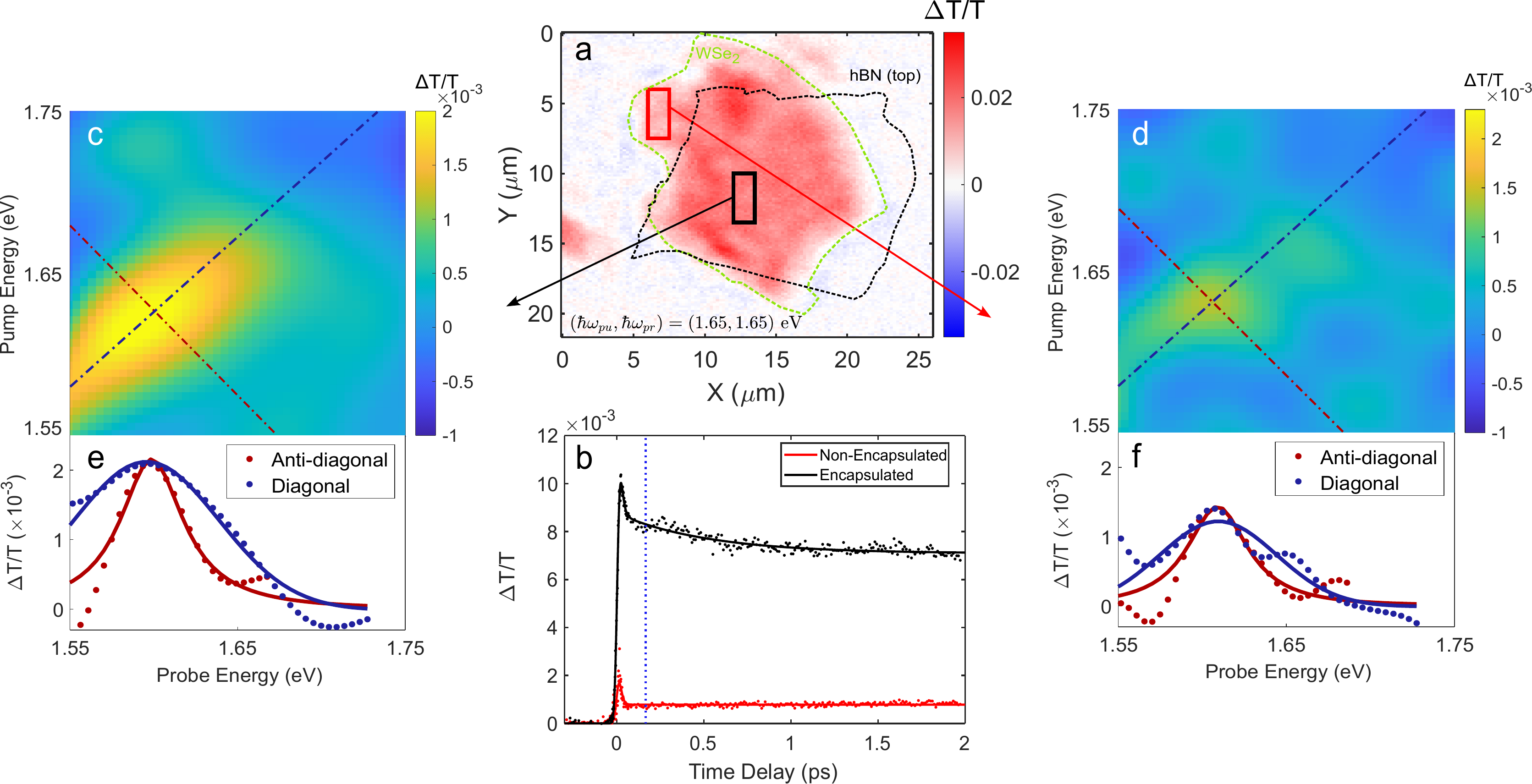}
\caption{\label{fig:results}\textbf{2DESM on bilayer WSe$_2$.} \textbf{(a)} \textbf{Spatially resolved signal} - The $S(x,y;\omega_{pu},\omega_{pr},t_{del})$ signal acquired at $t_{del}$=150 fs and for $\hbar \omega_{pu}$=$\hbar \omega_{pr}$=1.65 eV is shown. The transmission variation has been normalized by taking the ratio between the measured signal and the total space and frequency dependent transmission. The black and red areas correspond to fully encapsulated and unprotected bilayer WSe$_2$, as previously indicated in Fig. \ref{fig:sample}a. The contoured area in green highlights the WSe$_2$ region, while the black dashed contour marks the region encapsulated by hBN.
\textbf{(b)} \textbf{Time traces} - Relaxation dynamics of the $S(x,y;t_{del})$ signal as a function of the pump-probe delay $t_{del}$ obtained by integrating over the black and red areas and over the $\omega_{pu}$ and $\omega_{pr}$ frequency axes. The solid lines are the multi-exponential fit to the data.  \textbf{(c,d)} \textbf{2DES spectra} - 2DES $S(x,y;\omega_{pu},\omega_{pr},t_{del})$ signals for fixed $t_{del}$=150 fs and integrated over the black (c) and red (d) regions. The dashed blue lines represent the diagonal ($\omega$) and the dashed red lines represent anti-diagonal ($\tilde{\omega}$) directions, along which the lineshapes reported in panels (e,f) have been extracted. \textbf{(e,f)} \textbf{Lineshape analysis} - The panels report the diagonal (blue dots) and anti-diagonal (red dots) lineshapes obtain by taking the line profiles of 2DES spectra (see panels (c,d)) along the $\omega$ and $\tilde{\omega}$ directions. The red and blue curves represent the fit to Eqs. \ref{eq:S_diag} and \ref{eq:S_antidiag}, respectively.}
\end{figure*}
Figure \ref{fig:results} presents the results of the 2DESM measurements performed on bilayer WSe$_2$ at $T$=300 K. The excitation fluence is $\approx$30 $\mu J cm^{-2}$ distributed over a circular spot with $\sim$ 190 $\mu$m diameter (FWHM). The probe beam is spectrally identical to the pump and has $\approx$10 times lower intensity. 
The images are cropped down to a size of 140$\times$120 pixels and are acquired at a frequency of 3 kHz, with an integration time of 1 second. For each camera frame, whose acquisition is triggered by the chopper status (see scheme in Figure \ref{fig:setup}c), 6 laser pulses are emitted. The $S(x,y;\omega_{pu},\omega_{pr},t_{del})$ hypercube is thus obtained by averaging $N$=3000 normalized images (see Eqs. \ref{eq:T_pumped} and \ref{eq:T_unpumped}) in 1 second. For each $t_{del}$ delay, the $t_{pu}$ and $t_{pr}$ axes are scanned by acquiring 150 values corresponding to the birefringent wedge positions of both ICPG and ICS ranging from $–0.8$ to $1.2$ mm, corresponding to the $t_{pu}$ and $t_{pr}$ spanning  from $–55$ to $82$ fs. The typical acquisition time for the full data cube shown in Figure \ref{fig:2Ddata}a is 10 hours, depending on $N$ and the number of steps in both the ICS and ICPG. The frequency domain signal is obtained via FT (see Eq. \ref{eq:signal}) of the time-domain data apodized using symmetric sigmoid apodization window with 0.5 mm width
, chosen in order to maximize the SNR ratio and reduce spectral artifacts.

Figure \ref{fig:results}a shows the signal $S(x,y)$ extracted from the hypercube at fixed pump and probe frequencies ($\hbar \omega_{pu}$=$\hbar \omega_{pr}$=1.65 eV) and $t_{del}$=150 fs. Under these resonant excitation conditions, the signal is expected to be dominated by the A-exciton of bilayer WSe$_2$; however, since contributions from spectrally overlapping excitonic complexes such as trions cannot be excluded, we refrain from attributing the signal exclusively to the A-exciton. The green and black dashed lines indicate the WSe$_2$ bilayer and the hBN capping, respectively, as already discussed in Fig. \ref{fig:sample}a. Thanks to the capability of the technique to capture the spatial distribution of the transient transmissivity signal—measured simultaneously for all spatial locations within a single image—a spatial inhomogeneity of the signal across the sample is clearly visible. Here we will focus on the difference between the fully capped region (black rectangular area) and the region that is not protected by the hBN overlayer (red rectangular area). The colorscale highlights a lower signal from the unprotected area, as compared to the one from the center of the sample that is fully encapsulated by hBN. This difference can be further appreciated in panel (b) where we plot the signal $S(t_{del})$ at fixed pump and probe frequencies ($\hbar \omega_{pu}$=$\hbar \omega_{pr}$=1.65 eV), integrated over the black and red regions, as a function of the pump-probe delay $t_{del}$. The signal from the fully encapsulated region is almost 5 times larger and presents a double exponential dynamics characterized by a fast decay on the order of the pulse width and a slower decay on the picosecond timescale. The signal from the unprotected region, in contrast, after a similar fast dynamics, reaches a quasi-stationary value, which corresponds to a plateau within the temporal window probed in the experiment. The temporal dynamics of the two regions have been reproduced by using the following time-dependent function:
\begin{equation}
\label{eq:time}
f(t_{del}) = g(t_{del})*\left[H(t_{del})\cdot \left(\sum a_i e^{-t_{del}/\tau_i}+c\right)\right] 
\end{equation}
where $g(t_{del})$ is a Gaussian function accounting for the time resolution (see Sec. \ref{sec:resolution}), $*$ represents the convolution, $H(t_{del})$ is an Heaviside function, $a_i$ are the coefficients multiplying the exponential decays with time constants $\tau_i$ and $c$ is a constant accounting for the steady-state value at $t_{del}= 2 $ ps. The best fitting for the fully protected region is obtained by using $\tau_1=12\pm$4 fs and $\tau_2=0.5\pm$0.1 ps, whereas for the unprotected region the best fitting is obtained by using a single exponential decay with $\tau_1=8\pm$5 fs.
Additional insights on the exciton dynamics are given by the analysis of the 2DES maps shown in Figure \ref{fig:results}. Panels (c) and (d) show $S(\omega_{pu},\omega_{pr})$ for fixed time delay $t_{del}$=150 fs and integrated over the black (c) and red (d) areas. In the fully encapsulated area, the signal is centered at $1.598 \pm 0.002$~eV (extracted from diagonal line profile discussed below) and has an elliptical shape with a pronounced elongation along the diagonal axis, which is typical of systems in which disorder-driven inhomogeneous broadening is larger than the inverse of the intrinsic lifetime. 
Panel (e) reports the lineshapes along the diagonal ($\omega$) and antidiagonal ($\tilde{\omega}$) axes. Considering that in the partially collinear geometry \cite{Fuller2015} the 2DES signal measures the purely absorptive spectrum, which corresponds to the sum of the real part of rephasing and non-rephasing signals, the diagonal lineshape is described by the function \cite{Siemens2010,Milloch2024}:
\begin{equation}
\begin{aligned}
\label{eq:S_diag}
S_{\text{diag}}(\omega) = A \Bigg[ 
&\Re \Bigg\{ 
\frac{1}{\sigma (\gamma - i(\omega-\omega_0))} \,
\exp\!\Big( \frac{(\gamma - i(\omega-\omega_0))^2}{2\sigma^2} \Big) \\
&\times \Big[ 1 - \operatorname{erf}\!\Big( \frac{\gamma - i(\omega-\omega_0)}{\sqrt{2}\,\sigma} \Big) \Big]
\Bigg\} 
\Bigg]
\end{aligned}
\end{equation}
and the antidiagonal lineshape by \cite{Siemens2010,Milloch2024}:
\begin{equation}
\label{eq:S_antidiag}
S_{\text{antidiag}}(\tilde{\omega}) =
A \left[
\frac{1}{\gamma^2 + (\tilde{\omega}-\omega_0)^2}
\ast
\exp\!\left(-\frac{(\tilde{\omega}-\omega_0)^2}{2\sigma^2}\right)
\right]
\end{equation}
where $A$ is an amplitude coefficient, $\omega_0$ the central excitonic frequency, $\sigma$ is the disorder-induced inhomogeneous broadening and $\gamma$ is the intrinsic linewidth, whose inverse represents the actual lifetime of the exciton. The global fitting procedure applied to the diagonal and antidiagonal lineshapes returns $\gamma$=23$\pm$2 meV and $\sigma$=40$\pm$1 meV for the fully encapsulated WSe$_2$ bilayer. The inhomogeneous broadening is larger than the intrinsic linewidth, suggesting an important role played by the local disorder. The inverse intrinsic linewidth, i.e. $\hbar / \gamma\simeq29\pm 2$ fs, represents the actual decoherence time of the system, driven by the fastest scattering processes that can take place in the material. This short decoherence time is smaller than values reported in the literature in monolayer TMDs at low temperature \cite{Luo2023,Jakubczyk2019,Day2016,Cadiz2017}, as a consequence of the increased temperature-induced scattering with optical phonons \cite{Jakubczyk2018,Jakubczyk2016} and the additional linewidth broadening ubiquitously observed in bilayer as compared to monolayer samples \cite{raja2018enhancement}. These data suggest the rapid formation of an incoherent excitonic gas, which relaxes on the timescale $t_2$=0.5$\pm$0.1 ps (see Figure \ref{fig:results}b) due to the decay of $K-K$ direct excitons into $K-Q$ momentum-forbidden dark excitons \cite{Shi2023,Lindlau2018}, as observed by recent momentum-resolved photoemission experiment \cite{Madeo2020}.\\
In the unprotected region (red area in Fig. \ref{fig:results}a), we observe a suppression of the 2DES signal amplitude, which is centered at $1.611 \pm 0.003$~eV. The $13\pm5$~meV energy shift compared to the fully encapsulated region (black area in Fig. \ref{fig:results}a) is due to the different screening effects that influence the exciton photon energy, as already reported in the literature \cite{Cadiz2017, borghardt2017engineering, Gerber2018, Nerl2024}. Lineshape analysis (Fig. \ref{fig:results}f) similar to what has been done for the encapsulated region returns an intrinsic linewidth $\gamma$=22$\pm$3~meV, corresponding to $\hbar / \gamma\simeq30\pm 4$~fs decoherence time. The comparable intrinsic linewidth in the encapsulated and non-encapsulated regions suggests that, at room temperature, dephasing processes are dominated by scattering with thermally activated phonons \cite{Jakubczyk2018,Jakubczyk2016,Cadiz2017}, which are not affected by the hBN overlayer. 
\begin{figure*}[t] 
\centering
\includegraphics[width=1\linewidth]{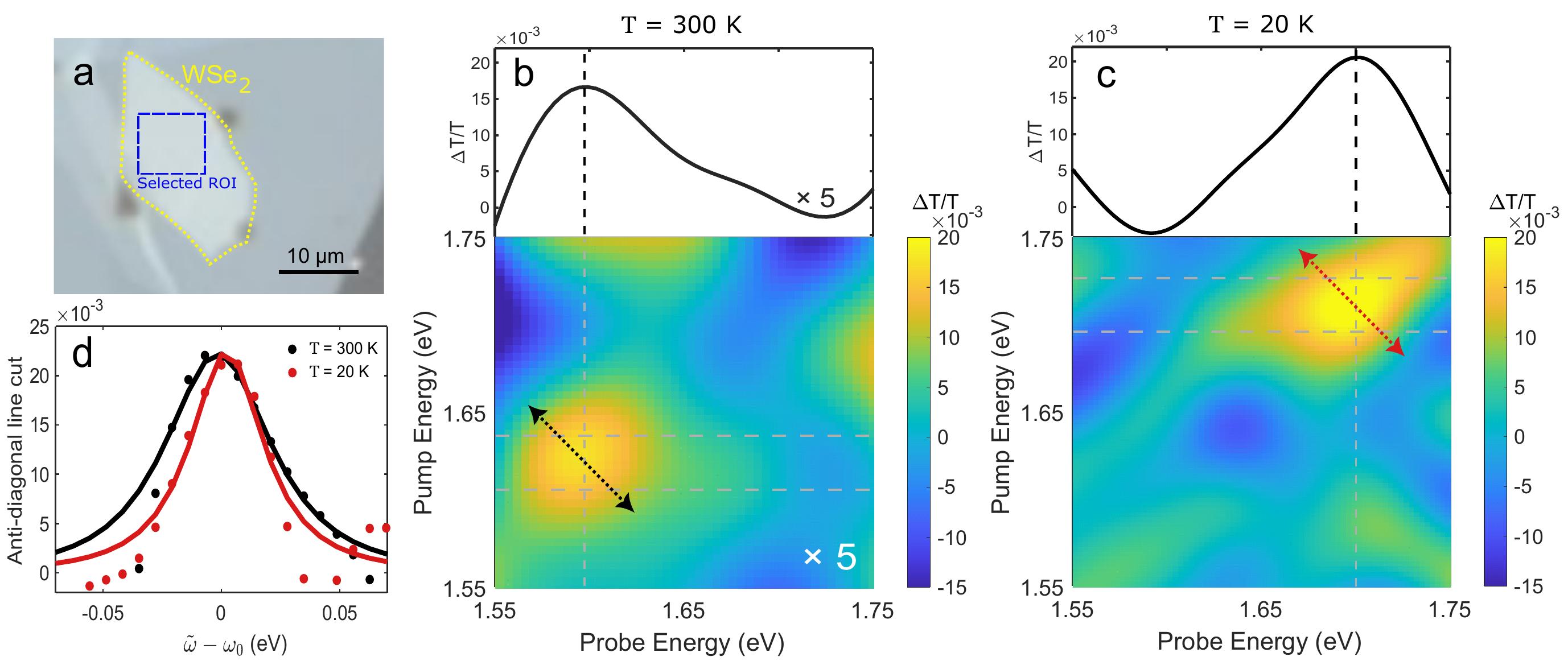}
\caption{\label{fig:Temp.results} 
\textbf{2DESM measurements on bulk WSe$_2$.}
(a) Optical microscopy image of the hBN-encapsulated sample. The yellow contour outlines the bulk WSe$_2$ region, while the blue rectangle indicates the area used for spatially averaging and extraction of the 2D maps. (b) 2DES map acquired at T=300 K and $t_{del}=50$ fs. The upper panel shows the pump-integrated signal obtained by integrating the 2D map over the pump-energy range defined by the horizontal gray dashed lines, revealing the excitonic resonance centered at 1.6 eV. For clarity, the signal amplitude is multiplied by a factor of five. (c) 2DES map acquired at T=20 K and $t_{del}=50$ fs. The upper panel shows the pump-integrated signal obtained by integrating the 2D map over the pump-energy range defined by the  horizontal gray dashed lines, with the excitonic resonance centered at 1.7 eV. (d) Antidiagonal line cuts at $t_{del}=50$ fs for both temperatures, together with fits along the $\tilde{\omega} $ direction. Solid lines correspond to fits using  Eq.~\ref{eq:S_antidiag}.}
\end{figure*}
\section{\textbf{Temperature-dependent 2DESM} }\label{sec.Temp}
To benchmark the 2DESM setup under cryogenic operating conditions, temperature-dependent measurements were performed on an hBN-encapsulated bulk WSe$_2$ flake. Figure \ref{fig:Temp.results}a shows an optical microscopy image of the employed sample, highlighted by the yellow contour. 2DESM measurements were carried out at $T$=300 K and $T$=20 K, for a fixed pump-probe delay t$_{del}$=50 fs, under the same experimental conditions of the data presented in Sec. \ref{sec:WSe2}, including excitation fluence, beam spot size, and image acquisition rate. At both temperatures, t$_{pu}$ and  t$_{pr}$ axes were scanned over the range $-$55 to 120 fs, yielding a spectral resolution improved by $\approx$1.5-fold, as compared to data presented in Sec. \ref{sec:WSe2}. \\
The 2DESM maps extracted from the central region of the flake (blue contour in Figure \ref{fig:Temp.results}a) are shown in Figures \ref{fig:Temp.results}b and \ref{fig:Temp.results}c for $T$ = 300 K and $T$ = 20 K, respectively. The low-temperature spectrum exhibits a fivefold enhancement of the excitonic resonance intensity, consistent with
the reduction of exciton–phonon scattering \cite{arora2015excitonic}. Concurrently, a blueshift by 90 meV is observed in the excitonic resonance energy, which moves from 1.600$\pm$0.005 eV at $T$=300 K to  1.693$\pm$0.003 eV at $T$=20 K, as highlighted by the horizontal line cuts displayed above each 2DES map, obtained by averaging the signal within the gray horizontal lines. This temperature-induced blueshift is in a good agreement with previously reported values for bulk WSe$_2$ \cite{arora2015excitonic}, and is attributed to bandgap renormalization driven by the progressive freezing of electron–phonon interactions and the consequent quenching of phonon-assisted sub-bandgap transitions upon cooling \cite{mishra2018exciton,ali2023temperature,arora2015excitonic}. \\
Beyond the spectral shift, temperature-dependent measurements reveal a pronounced evolution of the excitonic lineshape, reflecting substantial modifications of the exciton–phonon coupling strength and, consequently, of the intrinsic exciton decoherence time. The antidiagonal line cuts extracted at both temperatures are shown in Figure \ref{fig:Temp.results}d, where a clear temperature-driven narrowing of the antidiagonal linewidth is evident. Fitting Eq. \ref{eq:S_antidiag} to the experimental data yields a homogeneous linewidth that decreases from $\gamma$=25$\pm$6 meV at room temperature to $\gamma$=17$\pm$5 meV at 20 K, consistent with the low-temperature suppression of exciton–phonon scattering channels. This linewidth narrowing corresponds to an increase in the exciton decoherence time from $\tau=\hbar / \gamma\simeq26\pm6$ fs at room temperature to $\tau=\hbar / \gamma\simeq40\pm$10 fs  at 20 K. These observations demonstrate that, at high temperature, the exciton decoherence is accelerated by the coupling to the phonon thermal bath. Consistent with previous spatially integrated 2DES studies\cite{hao2016direct} and coherent spectroscopy investigations in transition metal dichalcogenides \cite{moody2015intrinsic}, the excitonic coherence is enhanced at low temperature due to reduced dephasing and approaches a low-temperature limit set by radiative recombination and disorder in the material. Overall, these findings demonstrate the capability of 2DESM to investigate excitonic coherence and dephasing processes as a function of temperature in quantum materials.

\section{conclusion}
In conclusion, we have developed a widefield two-dimensional electronic spectroscopy microscope (2DESM) that unifies multidimensional spectroscopy with optical imaging, enabling simultaneous femtosecond temporal, micrometer spatial, and broadband spectral resolution. This approach overcomes intrinsic limitations of pump–probe spectro-microscopy, which lacks sensitivity to quantum-coherent polarization dynamics and provides only population-based information. Compared to existing point-scanning 2DES microscopy approaches which require tight focusing with transmission objectives, 2DESM offers the key advantages of parallel spatial acquisition across the full field of view, enabling high-throughput hyperspectral imaging without the need for sequential raster scanning. Moreover, because no objective is employed in the excitation path, the technique supports a broader spectral bandwidth and higher temporal resolution. This comes at the cost of a moderate reduction in spatial resolution ($\sim1\mu m$) relative to diffraction-limited confocal implementations ($\sim0.3-0.5 \mu m$)\cite{Jakubczyk2016,Jakubczyk2018,Jakubczyk2019,liu2025nonlocal,Luo2023}. 
As a proof of concept, we applied 2DESM to bilayer WSe$_2$ encapsulated in hBN and resolved spatial variations in the A-exciton response. Encapsulated regions exhibited stronger 2DES signals, pronounced inhomogeneous broadening, and fast exciton decoherence ($\gamma$=23$\pm$2 meV), followed by a 0.5 ps relaxation into momentum-dark excitonic states. In contrast, uncapped regions showed reduced amplitude and a different relaxation pathway. Temperature-dependent measurements reveal a clear excitonic blueshift and enhanced coherence, associated with the low-temperature quenching of phonon fluctuations.  These results demonstrate the possibility of directly correlating local structural environment with quantum-coherent exciton dynamics, validating 2DESM as a general platform for interrogating disorder, dephasing, and energy-transfer processes in heterogeneous quantum-confined systems and for directly investigating these phenomena under cryogenic operating conditions.\\
2DESM, therefore, opens new opportunities for studying ultrafast quantum phenomena in real space, addressing open questions in excitonics and correlated material optoelectronics \cite{Regan2022} that remain challenging to investigate using pump–probe imaging or scanning-based 2DES. 2DESM has the capability to: i) resolve the intrinsic coherence limits at the microscale in van der Waals heterostructures, including systems with interlayer excitons and Moiré-trapped states; ii)
map the microscale landscape of disorder, allowing quantitative comparison of intrinsic linewidths versus inhomogeneous broadening across device architectures; iii)
study many-body interactions and emergent quasiparticles, such as biexcitons and trions, tracking where they form and how they propagate microscopically; iv) probe non-local coupling \cite{liu2025nonlocal} and energy transport, including coherent exciton diffusion and cavity-enhanced delocalization in photonic–TMD or semiconductor hybrids.\\
By unifying optical microscopy, quantum-coherence spectroscopy, and high-throughput hyperspectral imaging, 2DESM establishes a powerful experimental direction for quantum materials research, providing a novel prospective on the interplay between local structure, disorder, dielectric environment and the ultrafast electronic coherence in emerging nanoscale technologies.

\section*{Acknowledgments}
MA, RA, AM, FrP, FB and C.G. acknowledge financial support from MIUR through the PRIN 2020 (Prot. 2020JLZ52N-003) 
The authors also thank S. Pezzini for insightful discussions. 
\bibliographystyle{IEEEtran}
\bibliography{reference}
\end{document}